\documentclass[aps]{revtex4}

\usepackage{natbib}
\usepackage{color}
\usepackage[pdftex]{graphicx}
\usepackage{epsfig}
\usepackage{wrapfig}
\usepackage{sidecap}
\usepackage{subfig}
\usepackage{amsmath}
\usepackage{amssymb} 
\usepackage{fancyhdr}  
\numberwithin{equation}{section} 

\def\be{\begin{equation}}
\def\bea{\begin{eqnarray}}
\def\ee{\end{equation}}
\def\eea{\end{eqnarray}}
\def\bc{\begin{center}}
\def\ec{\end{center}}
\def\ba{\begin{array}}
\def\ea{\end{array}}
\def\bt{\begin{tabular}}
\def\et{\end{tabular}}
\def\bq{\begin{quote}}
\def\eq{\end{quote}}
\def\bi{\begin{itemize}}
\def\ei{\end{itemize}}

\def\1{{\pi}}
\def\v1{{\varpi}}

\def\v{\\[.1in]}

\def\'{{\prime}}
\def\bthm{\begin{thm}}
\def\ethm{\end{thm}}
\def\blm{\begin{lm}}
\def\elm{\end{lm}}
\def\bprop{\begin{prop}}
\def\eprop{\end{prop}}

\def\bs{\begin{slide}}
\def\es{\end{slide}}
\def\bbe{\begin{boldequation*}}
\def\ebe{\end{boldequation*}}

\begin{document}


\newtheorem{thm}{Theorem}
\newtheorem{lm}{Lemma}
\newtheorem{prop}{Proposition}
\title{Direct Numerical Simulation of Single-mode Rayleigh-Taylor Instability}
\author{Tie Wei}
\author{ Daniel Livescu}
\affiliation{Los Alamos National Laboratory, Los Alamos, NM, 87544 \\
        twei@lanl.gov, livescu@lanl.gov}

\maketitle
\vspace{-20pt}
Rayleigh-Taylor instability (RTI) is an interfacial instability that occurs when a high density fluid is accelerated or supported against gravity by a low density fluid. This instability is of fundamental importance in a multitude of applications, from fluidized beds, oceans and atmosphere, to inertial or magnetic confinement fusion, and to astrophysics. The interface between the two fluids is unstable to any perturbation with a wavelength larger than the cutoff due to surface tension (for the immiscible case) or mass diffusion (for the miscible case).  

The video shows the evolution of density and vorticity field from our Direct Numerical Simulation (DNS) of high perturbation Reynolds number single-mode RTI. The development of single-mode RTI can be divided into a number of stages, depending on which physical effect dominates the instability growth. At early times, if the initial perturbations amplitudes are small compared to their wavelength and the growth is not dominated by diffusive effects, the flow can be described by linearized equations and the perturbation amplitude grows exponentially with time (exponential growth stage-EG). With increasing bubble and spike speed, the differential velocity on the two sides of the interfaces leads to the development of the Kelvin-Helmholtz instability on the edges of the bubbles and spikes. However, not long after the non-linear effects become important, the vortical motions generated by the Kelvin-Helmholtz instability are weak, and the flow at the tip of the bubble is still potential.  This potential flow regime is characterized by a ``quasi-constant'' bubble front speed, and this staged is called `potential flow stage' (PFG). 

As the fluid accelerates due to the buoyancy forces, the initial vortices grow larger and start interacting. One of the first consequences of this interaction is that the vortices split and form pairs of counter-rotating vortices (one for each bubble and spike) which start self-propelling towards the tips of the bubbles and spikes. The motions become more complicated due to the further break-up, however, the first vortex pair still moves on an accelerating trajectory such that the induced velocity at the tips of the bubble/spike continues to increase. The consequence is that the velocity no longer follows the potential flow theory and the tips of the bubble/spike undergo a `re-acceleration stage'(RA). A new stage, chaotic development (CD), was revealed in our DNS after the re-acceleration stage. The chaotic development is caused by the complex vortical motions and interactions, which can be clearly in the later part of the movie. Since such complex motions have non-integrable dynamics, the bubble/spike velocities present chaotic temporal behavior.

The parameters used in the 2D simulation is shown in table \ref{tab:prms}.
\begin{table}[h]
\center
\begin{tabular}{|c|c|c|c|c|}
\hline
$L_h\times L_v$	&  $N_h \times N_v$	& $g$  & $\nu$ & $Sc$	\\
\hline
$2048\times10240$ & $2048\times12800$ & $11.0$ & $1.0$ & $1.0$ \\
\hline
\end{tabular}
\caption[]{Simulation parameters. $L_h$: domain size in the horizontal direction; $L_v$: domain size in the vertical direction; $N_h, N_v$: grid numbers in the horizontal and vertical, respectively; $g$: gravity; $\nu$: kinematic viscosity; $Sc$: Schmidt number.}
\label{tab:prms}
\end{table}

\end{document}